\documentstyle[multicol,epsfig,prl,aps]{revtex}

\begin{document}
\draft
\title{Premartensitic transition driven by magnetoelastic interaction in bcc
ferromagnetic Ni$_2$MnGa}
\author{Antoni Planes, Eduard Obrad\'o, Alfons Gonz\`alez-Comas, and Llu\'\i s Ma\~nosa}
\address{Departament d'Estructura i Constituents de la Mat\`eria. \\
Facultat de F\'\i sica. Universitat de Barcelona. Diagonal 647.\\
08028 Barcelona. Catalonia (Spain)}
\maketitle

\begin{abstract}
We show that the magnetoelastic coupling between the magnetization and the
amplitude of a short wavelength phonon enables the existence of a first
order premartensitic transition from a bcc to a micromodulated phase in Ni$%
_2 $MnGa. Such a magnetoelastic coupling has been experimentally evidenced
by AC susceptibility and ultrasonic measurements under applied magnetic
field. A latent heat around $9$ J/mol has been measured using a highly
sensitive calorimeter. This value is in very good agreement with the value
predicted by a proposed model.
\end{abstract}


\begin{multicols}{2}
\narrowtext

Martensitic transitions (MT) are first order displacive structural phase
transitions accompanied by a significant strain of the unit cell. In
general, homogeneous and intracell (short-wavelength phonons) strains are
necessary in order to describe the path followed by the atoms at the
transformation. An interesting feature displayed by the systems undergoing
this kind of first order transitions is the existence of precursor effects%
\cite{Krumhansl,Petry}. They reflect that, in a sense, the system prepares
for the phase transition before it actually takes place. For instance, in $%
bcc$ materials the $TA_2[110]$ phonon branch has low energy, the
corresponding elastic constant ($C^{\prime }$) is very low and softening of
all phonons and $C^{\prime }$ occurs on cooling \cite{Petry}.

The Ni$_2$MnGa Heusler alloy is investigated in the present work. At high
temperature it is ferromagnetic (the Curie temperature is $T_c=381$ K),
displays a $bcc$ structure with an $L2_1$ atomic order (space group $Fm3m$),
and transforms martensitically at $T_M=175$ K. This alloy is unique in the
sense that (i) it is the only known bcc ferromagnetic material undergoing a
MT and (ii) the MT is preceeded by a structural phase transition
(intermediate transition) to a micromodulated phase (the cubic symmetry is
preserved) resulting from the freezing of a $q=0.33$ $TA_2$ phonon which
becomes the intracell strain characterizing the new phase. Such a phase
transition has been evidenced by neutron scattering\cite{Zheludev95}, x-ray%
\cite{Fritsch}, electron microscopy \cite{Cesari97},and ultrasonic
measurements\cite{Worgull,Manosa}. We have recently suggested that this
transition to the premonitory (intermediate) phase is a consequence of the
magnetoelastic interplay between the phonon and the magnetization\cite
{Manosa}. At the MT the system transforms to a modulated structure with
tetragonal symmetry (homogeneous strain)\cite{Martynov}. The modulation of
the martensitic structure is different from that of the premartensitic
phase. It has been argued \cite{Manosa} that the intermediate transition has
to be first order because there is no complete softening of the frequency of
the soft phonon; nevertheless, attempts in measuring a latent heat have not
been successful \cite{Kokorin} and a small thermal hysteresis has only been
detected \cite{Zheludev96a} in samples subjected to external stresses (which
can result in a modification of the characteristics of the transition).

In this letter we present a phenomenological model for the intermediate
transition based on a Landau expansion, which includes a magnetoelastic
coupling. The primary order parameter is the amplitude $\eta $ of a TA$_2$ $%
[110]$ phonon, and secondary order parameters are: $\varepsilon $, a $(110)[1%
\overline{1}0]$ homogeneous shear suitable to describe a cubic to tetragonal
change of symmetry, and $M$, the magnetization (considered here to be a
scalar). In terms of these three order parameters we assume the free energy
function to have the following general form:

\begin{equation}
\label{freeenergy}{\cal F}(\eta ,\varepsilon ,M)=F_{str}(\eta ,\varepsilon
)+F_{mag}(M)+F_{me}(\eta ,\varepsilon ,M), 
\end{equation}
which includes a purely structural term $F_{str}$, a magnetic term $F_{mag}$%
, and a mixed term $F_{me}$ accounting for the magnetoelastic interplay.
Considering the symmetries of the system, the following expansions are
proposed for the three contributions: 
$$
\begin{array}{c}
F_{str}(\eta ,\varepsilon )= 
\frac 12m^{*}\omega ^2\eta ^2+\frac 14\beta \eta ^4+\frac 16\gamma \eta ^6+%
\frac 12c\varepsilon ^2 \\ F_{mag}(M)=- 
\frac 12AM^2+\frac 14BM^4\simeq A(M-M_0)^2-\frac 14\frac{A^2}B \\ 
F_{me}(\eta ,\varepsilon ,M)=\frac 12\kappa _1M^2\eta ^2+\frac 12\kappa
_2M^2\varepsilon ^2 
\end{array}
$$
All the coefficients in the above expansions are positive and only $\omega
^2 $ is supposed to be temperature dependent\cite{coefficients}. Actually $%
\omega $ is identified as the frequency of the anomalous phonon which
condensates at the intermediate transition. Experimentally, the square of
this frequency has been shown to exhibit a marked linear decrease on
approaching the intermediate transition \cite{Zheludev95}; hence we assume
that: $m^{*}\omega ^2=a(T-T_u)$. We have not included any direct coupling
between $\eta $ and $\varepsilon $, which is supposed to be negligible in
comparison with the magnetoelastic coupling. Concerning the purely magnetic
contribution $F_{mag}$, considering that the intermediate phase appears well
below the Curie point, the changes in the magnetization are expected to be
small; therefore, it is reasonable to linearize $F_{mag}$ around the value $%
M_0$ (the equilibrium magnetization close to the intermediate transition).

Minimization of ${\cal F}(\eta ,\varepsilon ,M)$ with respect to $%
\varepsilon $ and $M$ leads to an effective free energy function ${\cal F}%
_{eff}$ along a given transformation path ($\varepsilon =0$ and $M=M_0/[1+%
\frac{\kappa _1}{2A}\eta ^2]$) in terms of the phonon amplitude $\eta $.
Expanding the term $[1+\frac{\kappa _1}{2A}\eta ^2]^{-1}$ in power series
and keeping only the terms up to sixth order, we obtain:

\begin{equation}
\label{feffex}{\cal F}_{eff}=\frac 12m^{*}\widetilde{\omega }^2\eta ^2+\frac 
14\widetilde{\beta }\eta ^4+\frac 16\widetilde{\gamma }\eta ^6 
\end{equation}
where: $m^{*}\widetilde{\omega }^2=m^{*}\omega ^2+\kappa _1M_0^2=a(T-[T_u-%
\frac{\kappa _1M_0^2}a])=a(T-T_0)$ ; $\widetilde{\beta }=\beta -\frac{\kappa
_1^2M_0^2}A$, and $\widetilde{\gamma }=\gamma +\frac 34\frac{\kappa _1^3M_0^2%
}{A^2}$.

The interesting point to be stressed is: provided that $\frac{\kappa
_1^2M_0^2}A$ is large enough, $\widetilde{\beta }$ can become negative, and
in this case a first order transition can take place before the system
becomes linearly unstable at $T_0$($=T_u-\frac{\kappa _1M_0^2}a$). It is
also worth noting that, in case there was not magnetoelastic coupling, only
a continuous transition at $T=T_u$ would be possible. In general, a first
order transition is predicted by the Landau theory when the system
symmetries enable the existence of a cubic invariant in the free energy
expansion. However, in our case the possibility of the first order character
for the premartensitic transition is a consequence of the non-linear
coupling between the symmetry-breaking order parameter and the
magnetization. There is some experimental evidence that this transition has
a first order character. The main argument supporting this point is the fact
that the frequency of the anomalous phonon does not reach zero value at any
temperature\cite{Zheludev95}. Actually, $\omega ^2$ decreases linearly with
temperature and reaches a minimum (finite) value at the transition
temperature. Extrapolation down to $\omega =0$ predicts that a complete
phonon softening would occur $5$ K below the actual first order transition.
Based on this experimental evidence, in the following we will assume that in
Ni$_2$MnGa the premartensitic transition is first order, that is, $%
\widetilde{\beta }<0$.

The equilibrium first order transition temperature $T_{I\text{ }}$is
obtained from the two conditions: $\partial {\cal F}_{eff}/\partial \eta =0$
and ${\cal F}_{eff}(\eta _I)={\cal F}_{eff}(\eta =0)$, where $\eta _I=\pm (-3%
\widetilde{\beta }/4\widetilde{\gamma })^{\frac 12}$ is the value of the
order parameter of the distorted phase at the transition temperature. These
conditions lead to a transition temperature $T_I=\frac 3{16a}\frac{%
\widetilde{\beta }^2}{\widetilde{\gamma }}+T_0$, higher than $T_0$. The
corresponding entropy change at $T_I$ is obtained from:

\begin{equation}
\label{entchan}\Delta S=\left( \frac{\partial {\cal F}_{eff}}{\partial T}%
\right) _0-\left( \frac{\partial {\cal F}_{eff}}{\partial T}\right) _{\eta
_I}=-\frac 12a\eta _I^2=\frac{3a\widetilde{\beta }}{8\widetilde{\gamma }} 
\end{equation}
rendering a transition latent heat: $L=T_I\Delta S.$ It has been reported
that no latent heat has been detected using differential scanning
calorimetric techniques\cite{Kokorin}. Nevertheless, a heat capacity anomaly
(jump) has recently been observed at the premartensitic 
transition\cite{Manosa}. Such a jump can be calculated from our Landau 
model as:

\begin{equation}
\label{dec}\frac{\Delta C}{T_I}=\left[ \frac{\partial \Delta S}{\partial T}%
\right] _{T=T_I}=-\frac 12a\left( \frac{\partial \eta ^2}{\partial T}\right)
_{T=T_I}=-\frac{a^2}{\widetilde{\beta }} 
\end{equation}

From (\ref{entchan}) and (\ref{dec}) and using the expression of $T_I$, a
latent heat $L=2\left( T_0-T_I\right) \Delta C$ is obtained. $T_{0\text{, }%
}T_I$ and $\Delta C$ have been measured experimentally, thus enabling 
an\hfill evaluation\hfill of\hfill the\hfill latent\hfill heat. \hfill 
Taking\hfill $T_{0\text{ }}=225$\hfill K, 

\begin{figure}[tb]
\begin{center}
\epsfig{file=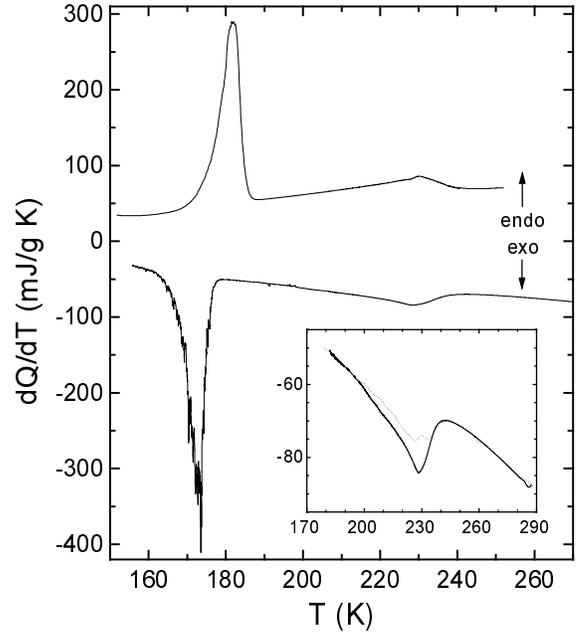, width=3.1in}
\caption{Thermograms recorded during cooling and heating calorimetric runs in a
Ni$_2$MnGa sample. The small peak corresponds to the premartensitic transition 
and the large one to the martensitic transition. The inset shows the premartensitic peak 
with the measured base line for evaluation of the latent heat.}
\label{FIG.1}
\end{center}
\end{figure}

\noindent $T_I=230$~K and 
$\Delta C\simeq 0.7$ J/mol K, the values $L\simeq -7$ J/mol and $\Delta
S\simeq -0.03$ J/mol K are obtained. These values are quite small (for
comparison note that at the MT the latent heat $L$ is around $100$ J/mol and 
$\Delta S\simeq 0.5$ J/mol K).

In order to corroborate the validity of the values predicted by the model,
calorimetric measurements have been carried out using a highly sensitive
scanning microcalorimeter\cite{Guenin}. This special calorimeter enabled the
use of large samples (this is not usual in standard differential scanning
calorimeters which are designed to operate with very small samples of a few
mg). A single crystal of Ni$_2$MnGa of $2.9$ g of mass and a Cu reference
with the same mass have been used. An example of the measured thermogram is
shown in Fig. 1. The large thermal effect corresponds to the MT. This
transition has a jerky character which is especially remarkable in the
exothermic forward transition. Also an intense acoustic emission has been
detected during this MT \cite{Gonzalez97}. A very small peak is clearly
observed above the MT. It corresponds to the premartensitic transition. The
dots shown in the inset below the peak correspond to a measurement of the
difference in the specific heats of a smaller sample of the same crystal\cite
{Composition} and a Cu reference, carried out using a modulated differential
scanning calorimeter. This measurement determines the base line that enables
a correct integration of the calorimetric peak leading to the transition
latent heat. We have obtained $L=$ $-9\pm 3$ J/mol. This\hfill value\hfill 
is\hfill in\hfill very\hfill good\hfill agreement\hfill with\hfill our\hfill 
numerical 

\begin{figure}[tb]
\begin{center}
\epsfig{file=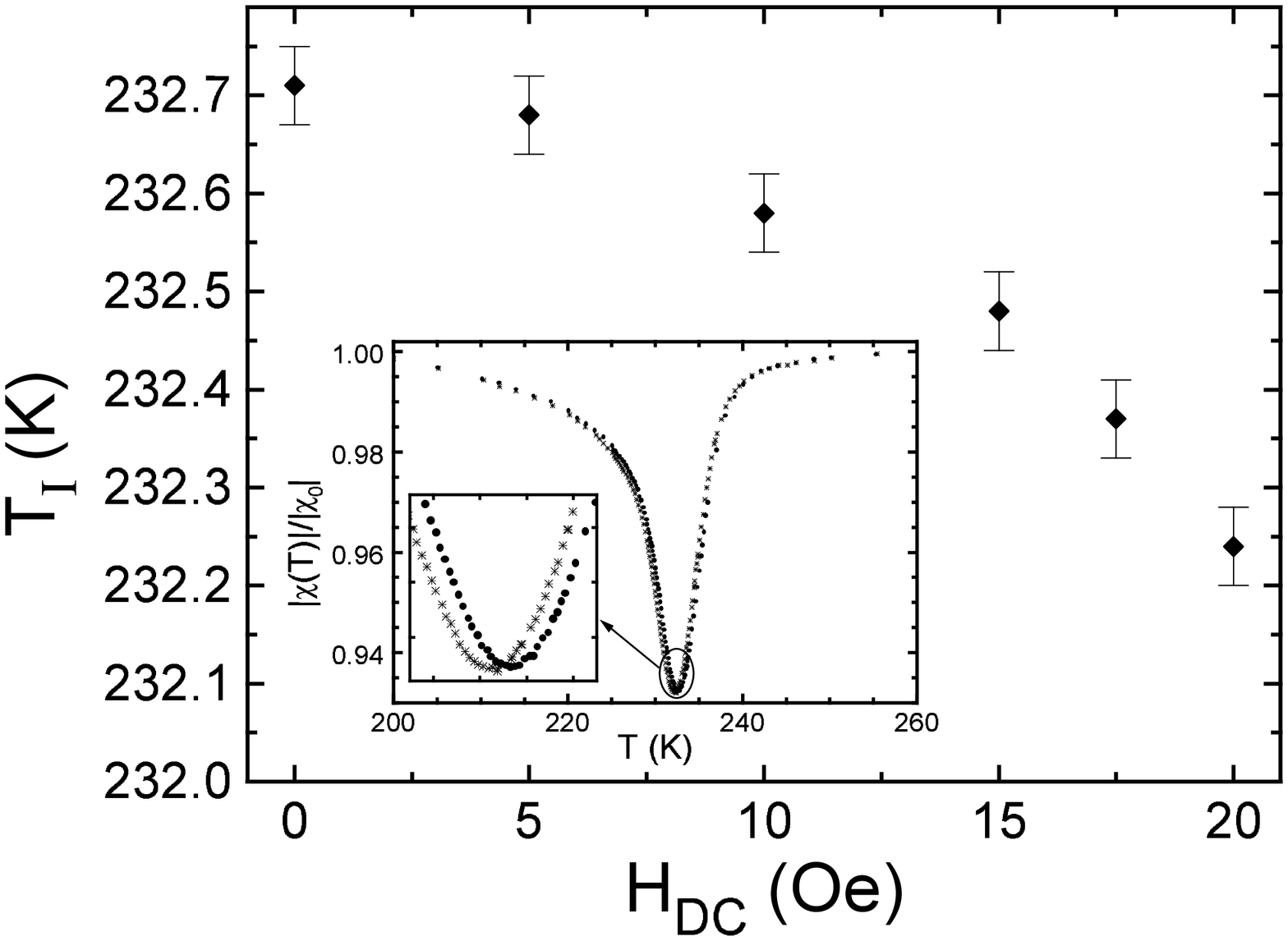, height=3.1in, angle=270}
\caption{Transition temperature as a function of a DC magnetic field applied
along the $[100]$ direction. The inset shows
an example of the AC magnetic susceptibility $\chi$ across the transition,
for two different applied fields (0 (dots) and 20 Oe (stars)). Data are normalized by
the measured value at 260 K ($\chi_0$). Measurements have been 
conducted with an AC field of 2 Oe and a frequency of 666 Hz .}
\label{FIG.2}
\end{center}
\end{figure}

\noindent prediction based on the Landau model.

For a first order transition, the Clausius-Clapeyron law must hold. In the
present investigation, the magnetoelastic interaction must lead to a change
in the transition temperature ($T_I$) under application of a magnetic field.
We have checked the field dependence of $T_I$ by measuring the magnetic
susceptibility accross the intermediate transition with different DC fields
applied along the $[100]$ direction, using an AC susceptometer. Results are
shown in Fig. 2. An unambiguous decrease in $T_I$ with increasing magnetic
field has been found. The change in $T_I$ becomes more pronounced at larger
values of the field. Such a non-linear behaviour seems to be due to a
decrease in $\Delta S$ as the field increases; this decrease would be
consistent with the fact that with the application of magnetic field, the
transition temperature approaches the unstability value. Moreover, the
observed decrease is in concordance with the behaviour obtained by computing 
$dT/dH$ from the model.

The elastic constant $C^{\prime }$ corresponding to a $(110)[1\overline{1}0]$
shear can also be obtained from the model as $C^{\prime }=c+\kappa _2M^2$.
With the aim of providing experimental evidence for this magnetoelastic 
effect we have measured the elastic constants under application of a
magnetic field at room temperature (ferromagnetic state) by the use of
ultrasonic techniques. These experiments will be reported in full elsewhere 
\cite{Gonzalez97}. In Fig. 3 we show an example of the behaviour exhibited
by the three independent elastic constants for a cubic symmetry. In all
cases, prior to each measurement, the sample was heated up to a temperature
well above the Curie point and cooled down to room temperature so that each
ultrasonic measurent corresponds to the first magnetization process. The
ultrasonic waves associated with $C^{\prime }$ were affected by a strong
scattering\hfill by\hfill the\hfill magnetic\hfill domains.\hfill This\hfill 
resulted\hfill in\hfill a\hfill larger 

\begin{figure}[tb]
\begin{center}
\epsfig{file=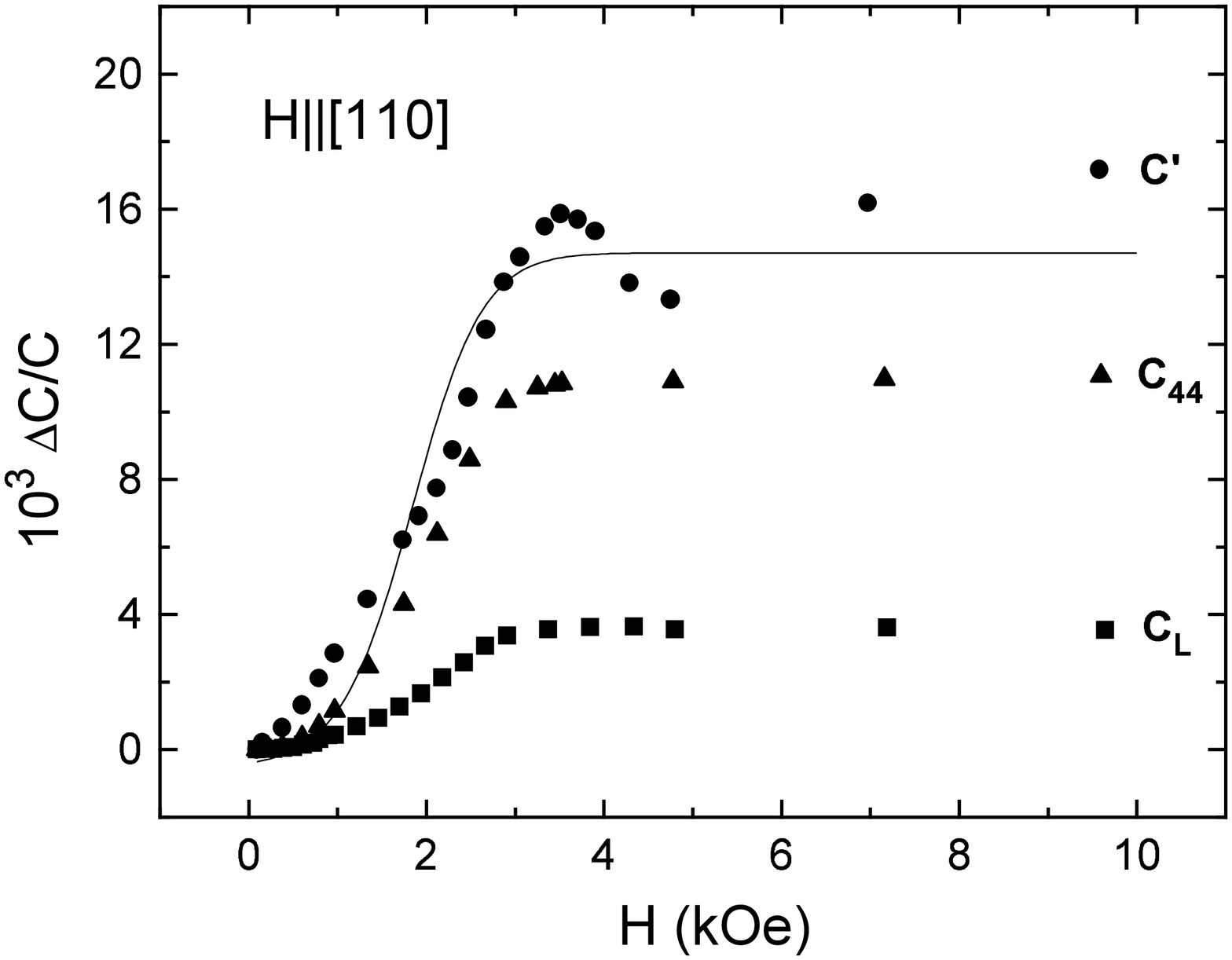, height=3.1in, angle=270}
\caption{Relative change of the elastic constants (solid symbols) as a function 
of an applied magnetic field $H$  
along the $[110]$ direcion. The solid line corresponds to 
$C'$ computed from measured values of $C_{11}$,$C_{44}$ and $C_{L}$. }
\label{FIG.3}
\end{center}
\end{figure}

\noindent error in the
determination of the field dependence of $C^{\prime }$. For this reason, we
also measured $C_{11}$, which enabled to compute $C^{\prime }$ as $%
C_{11}+C_{44}-C_L$; the result is shown as a continuous line in Fig. 3. We
have found that all elastic constants increase with the magnetic field up to
a saturation value\cite{CH}. For the same range of magnetic fields, an
increase of the magnetization from zero to a saturation value has been
reported in a similar sample \cite{Ullakko}; this indicates that the change
in the elastic constants presented here is associated to a change in the
value of the magnetization.

Across the intermediate phase transition, $M$ undergoes a change (we recall
that $M$ depends on $\eta $), which results in a jump of $C^{\prime }$ at
the transition, $\Delta C^{\prime }=\kappa _2(M_I^2-M_0^2)$, where $M_I$ is
the magnetization in the $\eta \neq 0$ modulated phase. Since $M_I<M_0,$ it
is expected that $\Delta C^{\prime }<0$. It is worth noting that this is in
agreement with measurements of the temperature dependence of $C^{\prime }$
which show a marked decrease of this elastic constant at the premartensitic
transition\cite{Worgull,Manosa}.

The model presented in this letter accounts for the magnetoelastic interplay
between the structural order parameters and the magnetization. Since in the
temperature range of interest Ni$_2$MnGa shows soft magnetic properties,
this interplay is mostly related to the reorientation of the magnetic
moments. The interaction between the magnetic and structural degrees of
freedom has been demonstrated by the measured field dependence of $C^{\prime
}$ and of the intermediate transition temperature. The coupling between the
amplitude of the anomalous short-wavelength phonon (primary order parameter)
and the magnetization gives the possibility for the occurrence of a first
order transition before a linear phonon unstability is reached. The
thermodynamic quantities of this transition have been obtained from the
model. They are in good agreement with experimental results. Actually, a
predicted latent heat of around $-7$ J/mol is in excellent agreement with
the value obtained from calorimetric measurements. The intermediate
transition takes place at a temperature slightly above the one where the
anomalous transverse phonon would become unstable. This indicates that the
first order character of this transition is weak, in concordance with the
small value found for the latent heat.

The microscopic origin of the phonon anomaly and magnetoelastic coupling
related to this phase transition are not explained by the proposed
phenomenological model. It has been suggested\cite{Zheludev96} that their
origin lies in the electron-phonon coupling and specific nesting properties
of the multiply connected Fermi surfaces. While this explains similar
short-wavelength phonon anomalies observed in Ni-Al alloys\cite{Zhao}, no
specific calculations in such a direction have, to our knowledge, been
performed in the case of the system studied here. Nevertheless, no
premartensitic transition has been reported to occur in Ni-Al. Actually, the
ferromagnetic character of the Ni$_2$MnGa is the main difference between
these two alloys.

The MT in non-ferromagnetic alloys have been described using Landau-type
models. In these models, the transition takes place as a consequence of the
anharmonic coupling between an anomalous transverse phonon and an
homogeneous strain\cite{Lindgard}. Such a coupling ($\eta ^2\varepsilon $)
is not included in our model. However, we argue that the magnetoelastic
interaction indirectly couples $\eta $ and $\varepsilon $. At the MT, $%
\varepsilon $ (tetragonal distortion) becomes different from zero and the
periodicity of the transverse modulation with wave vector $\zeta =0.33$ is
modified (a five layer modulation is obtained). To account for such a
modification in the periodicity, an explicit wavevector dependence of the
free energy should be included in the Landau model. It is also worth
noticing that reorientation of martensitic twin variants has been achieved
recently by the application of magnetic field \cite{Ullakko}.

In conclusion we have presented a model that accounts for the first order
phase transition between the $bcc$ and the intermediate phases. The first
order character has been experimentally demonstrated by the measurement of a
latent heat and by the magnetic field dependence of the transition
temperature. In the model, the first order transition occurs as a
consequence of a magnetoelastic coupling. Such a coupling has also been
experimentally evidenced. Hence, the premartensitic transition must be
considered as a magnetically driving precursor effect announcing the MT by
the modification of the dynamical response of the $bcc$ parent lattice.
Within a general framework, the results presented here evidence that
coupling of a secondary field (magnetic in this case) to incipient unstable
excitations, associated to structural degrees of freedom, can fundamentally
affect the characteristics of a phase transition.

This work has received financial support from the CICyT (Spain), Project No.
MAT95-504 and CIRIT (Catalonia), Project No. SGR00119. The single crystal
was kindly provided by V.A. Chernenko. E.O. and A.G. acknowledge financial
support from DGICyT (Spain).

\end{multicols}


\begin{references}
\bibitem{coefficients}  With the exception of $\omega $, all coefficients
are not expected to vary rapidly with temperature near the transition.
\bibitem{Composition}  The small sample has been cut from the single
crystal. It has been verified that the composition of the crystal is
homogeneous. Moreover, when possible, measurements have been conducted on
the two samples, and the obtained data have been found to be coincident
within the experimental errors.
\bibitem{CH}  The same behaviour is obtained with the magnetic field applied
along the [100] axis. In this case, for each elastic constant, the same
saturation value (within experimental error) is achieved at a lower value of
the magnetic field.
\end{references}
\end{document}